# COLLABORATION BETWEEN SAML FEDERATIONS AND OPENSTACK CLOUDS

MIHÁLY HÉDER

*Hungarian Academy of Sciences Institute for Computer Science and Control,*
*Kende u. 13-17, Budapest, 1111, Hungary*
*mihaly.heder@sztaki.mta.hu*

SZABOLCS TENCZER

*Hungarian Academy of Sciences Institute for Computer Science and Control,*
*Kende u. 13-17, Budapest, 1111, Hungary*
*tenczer.szabolcs@sztaki.mta.hu*

ANDREA BIANCINI

*Consortium GARR,*
*Via dei Tizii 6, Roma, 00185, Italy*
*andrea.biancini@garr.it*

In this paper, we present the design process of a novel solution for enabling the collaboration between OpenStack cloud systems in SAML federations with standalone attribute authorities, such as national research and education federations or eduGAIN. The software solution that realizes the integration of systems serves as a case study to show how abstract desirable engineering properties fixed at the beginning of the design process can be implemented during the development phase. An analysis of earlier generations of OpenStack-related developments trying to tackle the same problem is given. Many aspects of this software integration can be generalized to serve as a template for federative cloud access.

## 1. Introduction

SAML[1] identity and attribute federations are common in the research and education environment. They allow users to use their home institution credentials to access resources at partner institutions. This is achieved through the exchange of digitally signed XML[2] assertions. There are three major roles in mature SAML federations: service providers (SPs),[3] identity providers (IdPs),[4] and attribute authorities (AAs).[5] SPs provide resources for users. In this case study, The OpenStack[6] cloud is such a resource, with OpenStack as the SP. Technically, it is not necessary to have a one-to-one mapping between a resource and an SP. For example, it is possible to build a system in which the command line access and the web access of OpenStack represent two different SPs in SAML. The opposite is also possible, for instance, a set of independent web resources may be represented as one SP. IdPs are sources of user identity information that the SPs and AAs trust. Attribute authorities are sources of user attributes that SPs trust. User attributes include profile attributes (e.g. email, name, etc.) that users might control themselves, as well as authorization information managed by a community. Trust between these entities is pre-established through the exchange of signed metadata[7] that contains signing keys, trusted network endpoints and administrative information.

OpenStack is an open source, Infrastructure-as-a-Service (IAAS)[8] cloud system that is designed to be modular. The focus of this paper is on the authentication and authorization functionalities of the OpenStack system, which are designed to be highly configurable and extensible. Initially, OpenStack did not support SAML federations. Yet, thanks to its modular design, integration with SAML systems can be achieved in multiple ways. There have been a number of previous integration efforts (detailed in Section 2 Related Work) trying to resolve issues in this area. The two modules involved in the authentication and authorization process are Keystone,[9] OpenStack's authentication component, and Horizon,[10] the system's web interface.

### 1.1. *Desirable engineering properties*

The software solution presented here is both new and original. It was developed to achieve a certain set of *desirable engineering properties* lacked by earlier solutions. One of these was *encapsulation*[11] of new functionality within a self-contained module. This was especially important, because the alternative – implementing the functionality via source code patches to OpenStack – requires an update of the patches every time the host code changes. In the case of OpenStack's half-year release cycle, the necessary changes would have been frequent. Also, the merging of the functionality to one of the OpenStack mainline components would have undermined its modularity, and the OpenStack developers we contacted also advised against it.

*Reuse of mature components*[12] was also targeted for development. This mostly involved the reuse of SAML-related software components. The handling of SAML protocol and metadata requires complex logic, and since these components implement authentication, the correct design and implementation is critical to security. As a result, the development of any SAML-related code was avoided.

*Full compatibility* with SAML federations was set as a goal. This resulted in federated login, logout, the latest metadata defining the IdPs, metadata refresh, and the use of external attribute authorities, even multiple ones, in the same session. Compliance with legal requirements (such as the need of informed consent of attribute release) was also essential.

*Delegation of administration,*[13] in this case, user provisioning and authorization by external systems, was also achieved. SAML federations have a number of solutions for virtual organization and virtual group management[14] that can be relied on. All of the above properties helped contribute to our overarching goal of easy, long-term operation and maintenance.

## 2. Related Work

As previously mentioned, there have been many, mostly unrelated, SAML integration initiatives in OpenStack that are relevant to the current work.

(i) In 2012, David Chadwick from the University of Kent initiated a project to SAML-enable Keystone and OpenStack.[15,16] One drawback of this pioneering solution was that it did not include the reuse of existing middleware to handle the SAML protocol, but instead, it relied on SAML programming libraries to implement its own SAML functionality. As a result, it did not achieve full compatibility: it did not handle external attribute authorities, and it did not consume SAML metadata. This latter issue, however, leads to duplication of metadata. In this solution, OpenStack maintained a list of trusted IdPs in its own database format and did not rely on the federate metadata containing the same information in SAML XML.

(ii) In 2014, as a part of the HEXAA[17] project, Szabolcs Tenczer created a completely new solution[18] based on Shibboleth[19] as SAML middleware, and on OpenStack's ability to rely on external authentication modules. Andrea Biancini from GARR also contributed to this solution. The results were presented at OpenStack CEE Day 2015.[19] The main issue with this solution was that its source code was not encapsulated, but acted as a patch for the main OpenStack codebase, and therefore, would require future significant maintenance.

(iii) A completely new solution was included in the 2015-1 KILO[20] release of OpenStack. The primary assignee of this project at OpenStack was Adam Young from Red Hat, with additional contributors from CERN and IBM.[21] The approach was detailed at the OpenStack Cloud Identity Summit (slides 23–38).[22]

The solution is called WebSSO, a protocol-agnostic federation module that works with OpenID,[23] SAML, and other protocols. As WebSSO does not include SAML-related code, it makes it possible to encapsulate SAML functionality in a mature SAML middleware component. Shibboleth, mod_shib,[24] and the resulting Apache environment is used for authentication. In this example, the Keystone module is defended with Shibboleth (in solution (ii) it was Horizon, the web module). Using the WebSSO solution with Shibboleth achieves full compatibility, reuse of mature components and encapsulation. However, it is not able to create users, tenants and projects within OpenStack. Therefore, each user must first be created in OpenStack before s/he can login via WebSSO. As a result, the WebSSO solution does not fully achieve delegation of administration. The software solution described in this paper suggests a converged solution to overcome this, based on WebSSO.

(iv) There is also Keystone-to-Keystone SAML flow, supported by OpenStack. In this, Keystone acts as an IdP, and another Keystone instance as SP (see the OpenStack Cloud Identity Summit presentation (slide 40).[22] This solution is not compatible with SAML federations, as it does not consume SAML metadata, and also does not reuse mature components. However, it might be a viable solution for a completely different use case, in which an OpenStack user database is the identity source to be trusted by a federation.

### 3. Software Design

Our solution required a number of software design decisions to be made based on the design criteria outlined in the introduction. The most important was that OpenStack should be used in combination with TRL[25] 9 SAML middleware, so that proper handling of SAML-level actions were not demanded from OpenStack itself. This includes: (a) metadata handling, as per eduGAIN or other federation requirements, with signature verification, (b) handling of stand-alone AAs, (c) collaboration with discovery services, and (d) SAML single logout. As previously discussed, using OpenStack with mature SAML middleware achieved our goal of reusing mature components.

Moreover, it was important that the new software should not only be a patch to the OpenStack Horizon or Keystone components. Because of the *encapsulation* of new functionality in its own module, no regular patching of any other OpenStack components will be necessary. Python and Django were selected for consistency with other OpenStack components.

It was also important to align the software with the OpenStack Keystone project's vision for the future. At the OpenStack CEE Day 2015 event, a converged solution for the future was agreed with the head of the Keystone project, Morgan Fainberg. From the solution (ii), the user/project creation part would be separated into a standalone 'signup page' web application, using Keystone API calls, and the user/project/tenant created at signup, if necessary. The mainline code would then perform the rest of the authentication and map the SAML session to Keystone users. The rest of the (ii) solution would be discontinued, and Horizon would not be patched in the future. Fainberg indicated that the OpenStack Keystone project is open to accept improvement patches to (iii) as long as they do not involve user/tenant creation (or de-provisioning).

The solution needed to ensure that the user is always properly provisioned into Keystone before it makes contact with OpenStack. Otherwise, the user would successfully login into Shibboleth federation middleware, but would be denied access and greeted with an error message from Keystone. According to our design criteria, this should not be implemented in a hook of an OpenStack module. As a result, we relied on the sessionHook[26] ability of Shibboleth SP.

The software also needed to include the means to request user consent, achieved by a website presented to the user, if necessary. The resulting software, regsite, in collaboration with WebSSO (see Section 2, Related Work), implements the desirable engineering properties.

### 3.1. *Main workflow*

Figure 1 presents the main workflow of regsite. This workflow implements a collaboration between a SAML IdP, several SAML AAs, the SAML SP protecting Keystone, regsite, and finally, Horizon.

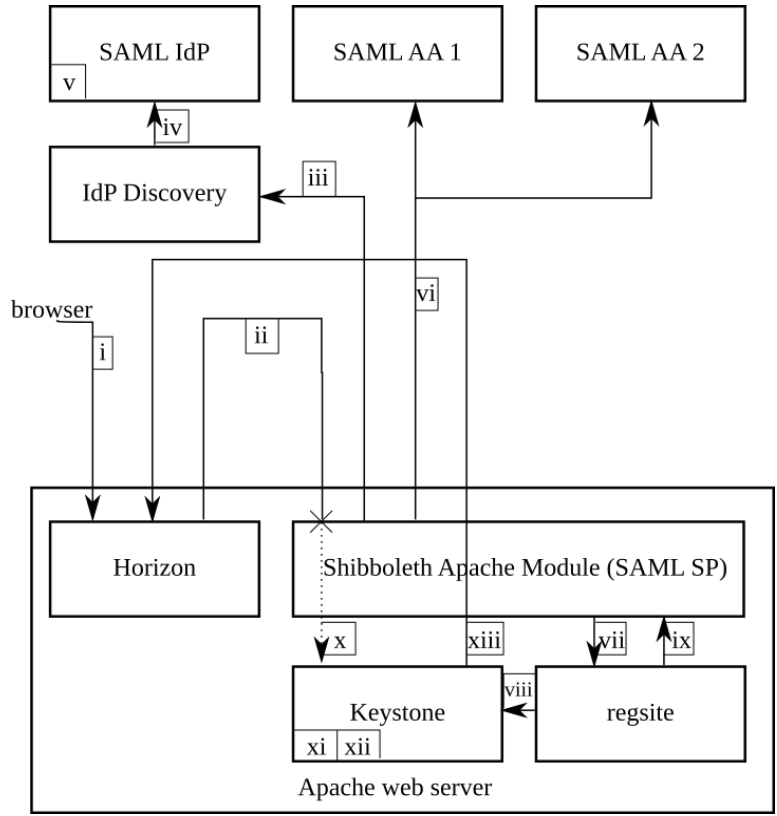

**Fig. 1:** The main workflow of the collaboration between SAML federations and OpenStack.

The workflow steps are as follows:
 (i) The user tries to access the OpenStack Horizon web interface with a web browser.
 (ii) Horizon redirects the user to the Keystone component's web endpoint.
 (iii) The Keystone component is hosted by an Apache web server and is guarded by a Shibboleth SP. The user does not have a Shibboleth session yet, therefore a SAML login sequence is initiated. The user forwarded to a SAML IdP discovery service, where s/he can select an identity provider.
 (iv) The discovery service forwards the user to the IdP.
 (v) The user logs in at the identity provider using his/her home institutional credentials.
 (vi) Additional profile attributes, and authoritative information is gathered from external attribute authorities, as defined by the SP's configuration. The number of AAs contacted can range from 0 to many, however, the SP sequentially queries the AAs, which aggregates the round-trip times of the single queries. Meanwhile, the user is blocked, which suggests that querying more than five AAs is not practical.

(vii) Shibboleth SP merges and filters the received attributes, then executes its configured sessionHook. It forwards the user to a location hosted on the same server as the SP, which also relays all the attributes gathered during the login process. In sessionHook, Shibboleth SP passes over the identity, profile and authoritative information to regsite. Steps (iii) to (vii) can all be completed by a standard Shibboleth SP.
(viii) regsite creates the user and the tenant, if necessary, using Keystone API calls.
(ix) regsite directs the user back into the Shibboleth login sequence.
(x) The Shibboleth login sequence finishes, and the user finally reaches Keystone. The same set of information is passed in Apache Environment variables to Keystone, as in Step (vii), to regsite.
(xi) Step (viii) ensures that the user is already existent in Keystone, as well as the tenants they are assigned to, therefore, Keystone successfully authenticates the user.
(xii) Keystone creates a token for the user.
(xiii) Keystone redirects the user to the Horizon web interface, accompanied by the newly created token. Horizon authenticates the user using this token and access is granted.

## 4. Implementation

The regsite implementation fulfils all the design requirements. It is a stand-alone Django[27] web application, designed to be run in an Apache web server and to be protected by Shibboleth. Although this possibility is outside the scope of the work discussed in this paper, theoretically, regsite should work with other SAML middleware and web server, and even in non-SAML scenarios.

The invocation of regsite into the login process is done by the Shibboleth sessionHook capability. The following XML start tag shows how the sessionHook is enabled in the *shibboleth2.xml* configuration.

```
<ApplicationDefaults
entityID="https://openstack.example.com/shibboleth"
sessionHook="/regsite">
```

A federate identifier, an entitlement (authoritative information), and an optional mail attribute from the Shibboleth middleware are received by regsite. The name of the actual attributes are configurable. A common setting is to use eduPersonPrincipalName[28] for the identifier, eduPersonEntitlement[29] or isMemberOf[30] for the entitlement, and mail[31] for the email address.

Mapping federate user identifiers to Keystone users is done by the following JSON configuration snippet. In this example, eduPersonPrincipalName (referenced in its abbreviated form: eppn) is used as a federate identifier. The *local* part of this configuration describes the Keystone user account (that is, local in relation to Keystone). The *remote* part identifies the federate user.

```
{
"mapping": {
```

```
        "rules": [
            {
                "local": [
                    {
                        "user": {
                            "domain": {
                                "id": "default"
                            },
                            "type": "local",
                            "name": "{0}"
                        }
                    }
                ],
                "remote": [
                    {
                        "type": "eppn"
                    }
                ]
            }
        ]
    }
}
```

The identifier received from the middleware will be used without modification by OpenStack. From the entitlement information, projects and roles are derived in the following way:
`<entilement_prefix>:project:role`

Divided by the colon (:), the segments are used to represent the project and role, which are OpenStack resources. There might be other colons in the entitlement prefix, but the software always uses the last two segments. Both the project and the role is created, as necessary. In case of multiple entitlement attributes (separated by semicolons), all of values are used. The logic of user, role and project creation by regsite is shown in Figure 2.

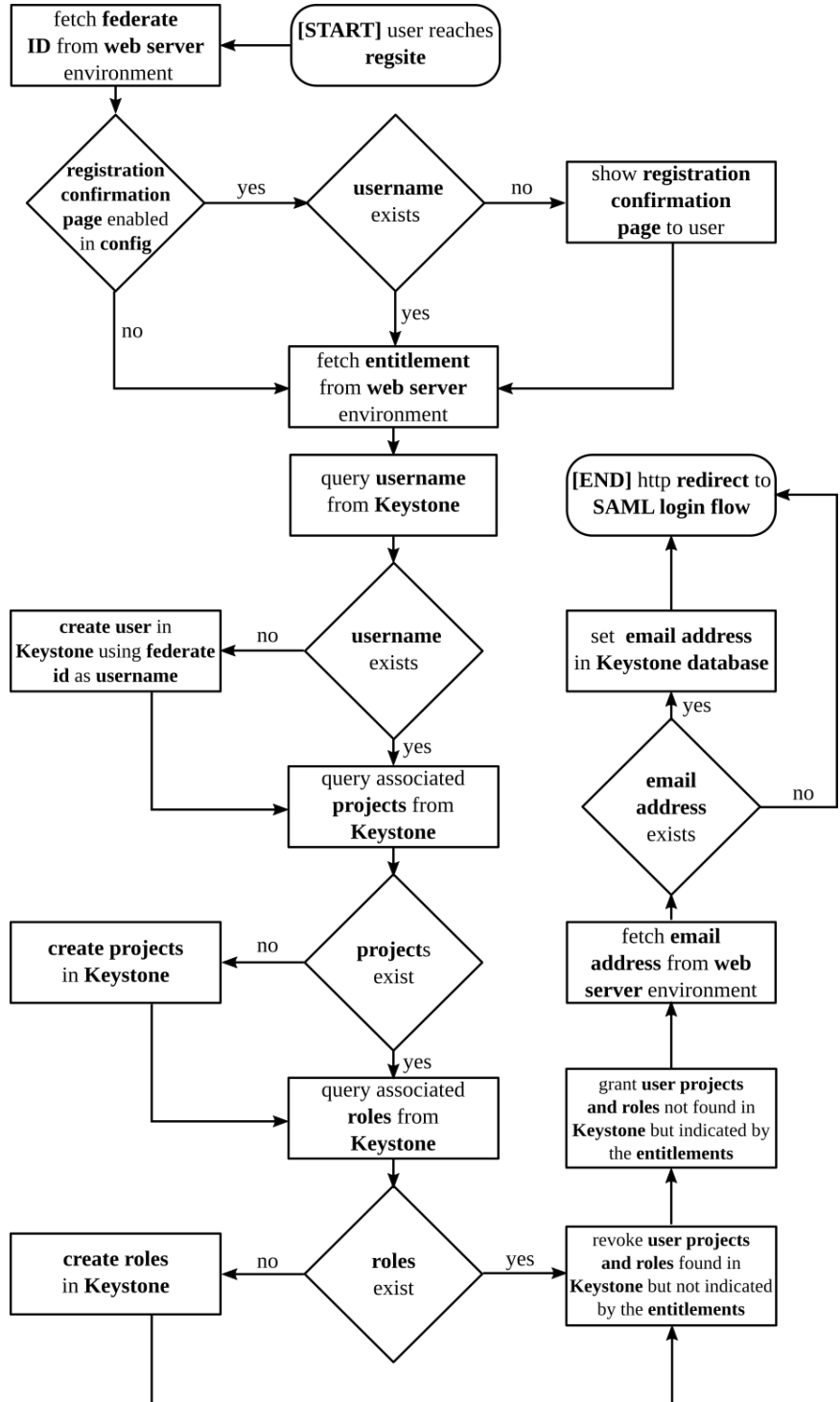

**Fig. 2** The flowchart of regsite user, role and project provisioning process

In the process shown in Figure 2, only the following, nine Keystone API calls are used by regsite to create users, roles and projects:

```
users.list, roles.list, projects.list, projects.create,
roles.create, users.create, roles.grant, roles.revoke,
users.update
```

As Keystone API is the only interface between regsite and Keystone, loose coupling is achieved. The *registration confirmation page* allows for requesting user consent before the actual user registration happens. At this point, the user can abandon the process before any data is relayed to OpenStack. This functionality ensures *legal compliance*. There are use cases in which the consent is acquired by other means, and also there are intra-organization use cases where consent is not necessary, as there is no new party involved. Therefore, this feature can be turned off by configuration.

## 5. Conclusion and Future Work

This paper presented a new method of collaboration between OpenStack cloud systems and mature SAML federations. Such collaboration was made possible by the modular design of OpenStack, which supports customized authentication and authorization, and by the generic nature of Shibboleth SAML middleware.

The discussed solution exhibits key engineering properties essential for long-term operation of deployed systems and maintenance of regsite code. Encapsulation of the new functionality necessary for the integration of OpenStack cloud systems and mature SAML federations is key for source code maintenance, while the reuse of mature components helps to minimize the size of that source code. Full compatibility with SAML ensures that the services provided by the federation – i.e. metadata distribution, discovery services, single login, and single logout – can be used. The fact that there is a very loose coupling between the systems, e.g. Shibboleth hides all SAML-related actions from OpenStack, and regsite relies on a small portion of Keystone API, ensures that the inevitable evolution of SAML federations and OpenStack will not endanger the easy maintenance of the collaboration.

However, this work is far from complete. Further research is needed in the area of user deprovisioning. Currently, new users of the federation with the correct permissions can access the resources in OpenStack, and their corresponding user accounts are created, as required, in the system. When users are no longer entitled to use the cloud, they will also be denied access to those resources. This is achieved by the collaboration of Shibboleth, regsite and OpenStack. However, the resources themselves will not be freed up, e.g. the virtual machines will not stop once a user's permissions have ceased. This is a very complex issue. First, there needs to be a policy in place to handle revoked rights. The immediate stop or deletion of resources that might include valuable data is probably unacceptable in most cases. A grace period and/or archival solution is necessary. Another challenge is the notification of the cloud system when a change is made to a user's entitlements in the SAML federation. This can follow a push or a pull model, for example, regsite could periodically pull the entitlements of known users in order to update OpenStack. Another solution is to implement an endpoint that can be notified by the group/virtual organization manager software of any changes.

Another area of future work relates to enabling command line access (CLI) while maintaining the properties of the integration, especially maintaining the reuse of mature components and loose coupling (that is, enabling CLI access without adding SAML-handling code to OpenStack). The issue stems from the fact that SAML is primarily designed for web resources.

One simple solution would be to extend regsite's user-facing web interface to provide a token for the user to use on CLI. SAML protects regsite, where users are both authenticated and authorized, so the proper token could be requested from Keystone on users' behalf. The users would then need to copy the token from the web and use them in the command line client. Alternatively, users could set up their own passwords in regsite for use on CLI, (although this password could not be used for web access, which is handled by Shibboleth). A more complex solution would involve relying on ECP. However, not every SAML IdP is ECP-enabled. Discovery and the querying of attribute authorities are also unresolved in this setup.

## Acknowledgements

The research leading to these results has received funding from the European Union's Horizon 2020 research and innovation program under Grant Agreement No. 691567 (GN4-1).

The authors are deeply indebted to the whole GN4-1 Joint Research Activity 3 (JRA3) Team, especially: Kristóf Bajnok, Maarten Kremers, Alejandro Perez Mendez, Remco Poortinga – van Wijnen and Michal Procházka. Many thanks also to Morgan Fainberg, for his valuable guidance in OpenStack Identity Issues.

## References

1. OASIS Security Services Technical Committee, Security Assertion Markup Language (2005), http://docs.oasis-open.org/security/saml/v2.0/saml-2.0-os.zip.
2. M. Bartel, J. Boyer, B. Fox, B. LaMacchia and E. Simon, XML Signature Syntax and Processing (Second Edition, 2008), http://www.w3.org/TR/xmldsig-core/.
3. OASIS Security Services Technical Committee, Glossary for the OASIS Security Assertion Markup Language (SAML) v 2.0 (2005), http://docs.oasis-open.org/security/saml/v2.0/saml-glossary-2.0-os.pdf (line 361).
4. OASIS Security Services Technical Committee, Glossary for the OASIS Security Assertion Markup Language (SAML) v 2.0 (2005), http://docs.oasis-open.org/security/saml/v2.0/saml-glossary-2.0-os.pdf (line 201).
5. OASIS Security Services Technical Committee, Glossary for the OASIS Security Assertion Markup Language (SAML) v 2.0 (2005), http://docs.oasis-open.org/security/saml/v2.0/saml-glossary-2.0-os.pdf (line 146).

6. The OpenStack Foundation, Open source software for creating private and public clouds (2015), http://docs.openstack.org/.
7. OASIS Security Services Technical Committee, Metadata for the OASIS Security Assertion Markup Language (SAML) v 2.0 (2005), http://docs.oasis-open.org/security/saml/v2.0/saml-metadata-2.0-os.pdf (line 1158).
8. M. Ali, S. Khan, and A.Vasilakos, Security in cloud computing: Opportunities and challenges, *Information Sciences* **305** (2015) 357–383, doi: http://dx.doi.org/10.1016/j.ins.2015.01.025 p. 360.
9. The OpenStack Foundation, OpenStack Identity (2015), http://docs.openstack.org/admin-guide-cloud/common/get_started_identity.html.
10. The OpenStack Foundation, OpenStack Dashboard (2015), http://docs.openstack.org/admin-guide-cloud/common/get_started_dashboard.html.
11. B. Meyer, *Object-Oriented Software Construction, Second Edition* (Prentice Hall, New Jersey, 1997), p. 53.
12. B. Meyer, *Object-Oriented Software Construction, Second Edition* (Prentice Hall, New Jersey, 1997), p. 67.
13. K. Bajnok, M. Héder, Z. Magyar and I. Tétényi: "HEXAA: Higher education external attribute authority" *Connect Magazine* **18** 14–15, http://issuu.com/danteprm/docs/connect_issue_18_web/17?e=6131560/11460567.
14. I. Tétényi, M. Héder, Z. Magyar, K. Bajnok, Open Call Deliverable OCK-DS1.1 Final Report (HEXAA), GÉANT (2013), p. 5. http://www.geant.net/Resources/Open_Call_deliverables/Documents/HEXAA_final_report.pdf
15. D. Germonville, Y. Fouillat and D. Chadwick, Adding federated access to OpenStack (2012),
https://dl.dropboxusercontent.com/u/44986510/Adding%20federated%20access%20to%20OpenStack%201.pdf.
16. D. Chadwick, Adding Federated Identity Management to OpenStack, *The OpenStack Summit* (2012), https://www.openstack.org/summit/san-diego-2012/openstack-summit-sessions/presentation/adding-federated-identity-management-to-openstack.
17. Hungarian Academy of Sciences Institute for Computer Science and Control and National Information Infrastructure Development Institute, HEXAA project site (2014), https://sites.google.com/a/sztaki.hu/hexaa/.
18. S. Tenczer, Shibboleth authentication backend for Horizon (2015), https://github.com/burgosz/openstack-horizon-shibboleth.
19. S. Tenczer, M. Héder, OpenStack SAML Integration with HEXAA, *OpenStack CEE Day* (2015), http://openstack.hexaa.eu/.
20. The OpenStack Foundation, OpenStack 2015.1.0 (Kilo) Release Notes (2015), https://wiki.openstack.org/wiki/ReleaseNotes/Kilo.
21. A. Young, S. Martinelli, M. Denis, J.C. Leon and T. Tran, Web Single Sign On Portal (2015): http://specs.openstack.org/openstack/keystone-specs/specs/kilo/websso-portal.html.
22. S. Martinelli, Building IAM for OpenStack, *OpenStack Cloud Identity Summit* (2015), http://www.slideshare.net/SteveMartinelli1/building-iam-for-openstack.

23. R. David, and D. Reed, OpenID 2.0: a platform for user-centric identity management. *Proceedings of the second ACM workshop on Digital identity* management. ACM, (2006).
24. S. Cantor, S. Carmody, M. Erdos, K.Hazelton, W. Hoehn, R.L.B Morgan, T. Scavo, D. Wasley, Shibboleth Architecture (2005), https://wiki.shibboleth.net/confluence/download/attachments/2162702/internet2-mace-shibboleth-arch-protocols-200509.pdf.
25. Horizon 2020 Work Programme 2014–2015, Technology Readiness Levels (2013), http://ec.europa.eu/research/participants/data/ref/h2020/wp/2014_2015/annexes/h2020-wp1415-annex-g-trl_en.pdf.
26. Native Shibboleth SP Application (2005), https://wiki.shibboleth.net/confluence/display/SHIB2/NativeSPApplication.
27. Django Software Foundation, The Web framework for perfectionists with deadlines (2015), https://www.djangoproject.com/.
28. Internet2 Middleware Architecture Committee for Education, Directory Working Group, eduPerson Object Class Specification (2012), http://www.internet2.edu/media/medialibrary/2013/09/04/internet2-mace-dir-eduperson-201203.html#eduPersonPrincipalName.
29. Internet2 Middleware Architecture Committee for Education, Directory Working Group, eduPerson Object Class Specification (2012), http://www.internet2.edu/media/medialibrary/2013/09/04/internet2-mace-dir-eduperson-201203.html#eduPersonEntitlement.
30. K. Hazelton, Internet2 Middleware Initiative, LDAP representations of memberships and groups (2015), http://macedir.org/specs/internet2-mace-dir-ldap-group-membership-200507.html.
31. Internet2 Middleware Architecture Committee for Education, Directory Working Group, eduPerson Object Class Specification (2012), http://www.internet2.edu/media/medialibrary/2013/09/04/internet2-mace-dir-eduperson-201203.html#mail.